\documentclass[11pt]{article}
\usepackage{graphicx}
\usepackage{latexsym}
\usepackage{amsfonts}
\newtheorem{lemma}{Lemma}
\newcommand{\E}{\mathbb{E}}
\begin{document}

\title{\`A la Carte of Correlation Models: Which One to Choose?}
\author{
Harry Zheng\thanks{Department of Mathematics,
Imperial College, London SW7 2BZ, UK. h.zheng@imperial.ac.uk.}\\
Imperial College}
\date{}
\maketitle 

\noindent
{\bf Abstract.} \  
In this paper we propose a copula contagion mixture model 
for correlated default times. The model includes the well known 
factor, copula, and contagion models as its special cases. The key advantage of such a model is that we can study the interaction of different models and their  pricing 
impact.  Specifically, we model the marginal default times to follow some  contagion intensity processes coupled with 
 copula dependence structure. We apply the total hazard construction method to generate
ordered default times and numerically compare the pricing impact of different
models on basket CDSs and CDOs in the presence of exponential decay and counterparty
risk. 



\section{Introduction}
The recent financial crisis has profound impact on the financial systems in the US, UK, and other major markets. Some giant banks and insurance companies either collapsed or had to be bailed out by the national governments.  The excessive risk 
exposure of many banks to collateral debt obligations (CDOs)
and credit default swaps (CDSs) has played the key role in this financial
crisis. One may list many causes which have attributed to
and aggravated the crisis, however, in this paper we only focus on the impact
of correlation modelling on the pricing of these portfolio credit derivatives. 

CDOs and CDSs had phenomenal growth in recent years until this financial
crisis. The key in 
pricing and hedging these portfolio credit derivatives is the correlation modelling.
 There are  mainly three approaches in the literature:
conditional independence, copula,  and  contagion. 
Factor models are  most popular due to their semi-analytic 
tractability. 
Many effective algorithms have
been developed to characterize the portfolio loss distribution, 
see Andersen et al.~(2003), Hull and White (2004) for recursive exact methods, 
and Glasserman (2004), Zheng (2006) for analytic approximation methods.
Factor models  may underestimate the
portfolio tail risk and economic capital,  see Das et al.~(2007).
Copula models are also popular, especially  
the Gaussian copula which is used in CreditMetrics,
see Li (1999).  Some copulas (Archemedian and exponential)
are good to model extreme tail events and simultaneous defaults. 
There are some active recent debates on the usefulness of copulas in financial modelling and risk management, see Mikosch (2006) and many  discussion papers on the same issue. 
Contagion models study the direct interaction of names in which
the default intensity of one name may change  
upon defaults of other names
and  ``infectious defaults'' may develop, see Davis and
Lo (2001), Jarrow and Yu (2001), Yu (2007). 
It is in general difficult to characterize the  joint distribution of
 default times 
 due to the looping dependence structure. 
For homogeneous portfolios  there is a closed form formula
for the density function of ordered default times, see Zheng and
Jiang (2009).
Monte Carlo method is often used to price
CDOs and basket CDSs no matter which correlation model is used
and provides benchmark results to test efficiency and
accuracy of analytic and numerical algorithms.  

It is interesting to know 
which model one should choose in pricing portfolio credit derivatives. 
We know different models give different values. If one uses the Gaussian copula, the swap rate for senior tranche of CDO is low due to the
thin tail distribution of portfolio loss, on the other hand, if one uses the contagion model, the swap rate for the same senior tranche is much higher.  However, one cannot simply say the contagion model is preferable to the Gaussian copula because it provides higher swap rates for senior tranche. It all depends on the underlying model assumptions. These correlation models are defined under different frameworks and are difficult to compare directly their pricing impact. It is therefore beneficial to have a
unified model which covers all 
three known models as special cases. One may then 
extract the information of the interaction of these models and may give a more balanced view on which model one should choose for a specific application.

In this paper we  
suggest a general copula contagion mixture model which includes factor, copula, and contagion models as its special cases. The key advantage of such a model is that we can study the interaction of different models and their  pricing 
impact.  Specifically, we model the marginal default times to follow some exponential decay contagion intensity processes coupled with some copula dependence structure. This is not a Markov process model and cannot be solved with the standard Kolmogorov equations or matrix exponentials,
see  Frey and Backhaus (2004),
 Herbertsson and Rootzen (2006). Although there are analytic pricing formulas for some special cases, we choose to use the Monte Carlo method to price CDOs and basket CDSs, which is reliable, accurate and efficient with some optimized numerical procedure. 

The paper is organized as follows:
section~2 describes the copula contagion mixture model 
and the relation with the known models,
 section~3 applies the model to price CDOs and basket CDSs and discuss the impact of interaction of different models, 
section~4  concludes.

\section{The Model}
Let $(\Omega,{\cal F}, \{{\cal F}_t\}_{t\geq 0}, P)$ be a
filtered probability space, where $P$ is the martingale
measure and $\{{\cal F}\}_{t\geq 0}$ is the filtration satisfying the usual
conditions.
Let $\tau_i $ be the default time of name $i$,  $N_i(t)=1_{\{\tau_i\leq
t\}}$ the default indicator process of name $i$, 
${\cal F}^i_t=\sigma(N_i(s): s\leq t)$ the filtration generated by default
process $N_i$, $i=1,\ldots,n$, 
and ${\cal F}_t= {\cal F}^1_t \vee \ldots
\vee {\cal F}^n_t$
the smallest $\sigma$-algebra needed to support $\tau_1,\ldots,\tau_n$.
Assume that $\tau_i$ possesses a nonnegative ${\cal F}_t$ predictable intensity
process $\lambda_i(t)$ satisfying $\E[\int_0^t \lambda(s)ds]<\infty$ 
for all $t$, and the compensated process 
$$ M_i(t)=N_i(t)-\int_0^{t\wedge \tau_i} \lambda_i(s)ds$$
is an ${\cal F}_t$ martingale.  
 Given $\tau_{j}=t_{j}$,
$j\in J_k=\{j_1,\ldots,j_k\}\subset \{1,\ldots,n\}$, 
satisfying $0=t_{j_0}<t_{j_1}<\ldots<t_{j_k}$ and $\tau_i>t>t_{j_k}$ for
$i\not\in J_k$, the  conditional hazard rate
of $\tau_i$  at time $t$ is given by
$$ \lambda_i(t|t_{J_k})=
\lim_{\Delta t\downarrow 0} {1\over \Delta t}
P(t<\tau_i\leq t+\Delta t| \tau_j=t_j, j\in J_k)$$
where $t_{J_k}$ is a short form for $(t_{j_1},\ldots,t_{j_k})$.

In copula modelling of default times it is normally assumed that
intensity processes are independent of default times of 
other names, i.e., $\lambda_i(t|t_{J_k})=\lambda_i(t)$ 
for all $t$. The marginal distribution functions of default times
$\tau_i$ are given by $F_i(t)=P(\tau_i\leq t)=
\E[1-\exp(-\int_0^t \lambda_i(s)ds)]$ 
(if $\lambda_i$ are stochastic processes) 
and standard uniform variables 
$F_i(\tau_i)$, $i=1,\ldots,n$, have a joint distribution function $C$, a given copula. 
It is easy to generate default times $\tau_i$ with the Monte Carlo method.
One can simply first generate standard uniform variables $U_i$
 with copula $C$ and 
sample paths of $\lambda_i$  and then 
find the default times $\tau_i$ by
$$ \tau_i = \inf\left\{t>0: \int_0^t \lambda_i(s)ds \geq E_i\right\}$$
where $E_i=\ln(1-U_i)$, $i=1,\ldots,n$, are correlated standard exponential
variables. In particular, if $\lambda_i(t)=a_i$, then $\tau_i=E_i/a_i$. 

There has been extensive research in literature on factor modelling of default times. These models are all special cases of copula modelling of default times. For example, the well-known Gaussian factor model is given by
$$ X_i=\rho Z+\sqrt{1-\rho^2}Z_i$$
where $Z,Z_1,\ldots,Z_n$ are independent standard normal variables
and $\rho$ is a constant satisfying $|\rho|\leq 1$. 
$Z$ is often interpreted as a systematic  factor and $Z_i$ idiosyncratic factors. If we set $U_i=\Phi(X_i)$, where $\Phi$ is the standard normal distribution function, then the distribution of $U_i$ is a Gaussian copula given by
$$C(u_1,\ldots,u_n)=\Phi_{m,\Gamma}(\Phi^{-1}(u_1), \ldots,
\Phi^{-1}(u_n))$$
 where $\Phi_{m,\Gamma}$ is the $n$-variate
standard normal distribution function  with mean vector $m=0$ and correlation matrix $\Gamma$ that has diagonal elements 1 and all 
other elements $\rho^2$. 
Factor models are appealing from model interpretation 
and conditional independence point of view. The corresponding copulas may have some complex forms, but in general it is easy to generate correlated standard uniform variables due to the special structure of factor models. From mathematics point of view  there is no need to 
treat them separately if we know how to generate standard uniform variables
$U_i$ from given copulas $C$.

In contagion modelling of default times 
the intensity processes 
$\lambda_i(t|t_{J_k})$ depend on default times of other names
and standard uniform variables $F_i(\tau_i)$ are assumed to be 
independent.  The marginal distribution functions $F_i$ 
of default times $\tau_i$ cannot be simply expressed in terms of $\lambda_i(s)$ as information of other default times is needed to characterize the whole intensity process paths and there is a ``looping'' phenomenon.
Although it is difficult to characterize the marginal and 
joint distributions of default times it is easy and straightforward to generate ordered default times $\tau^i$
 with the
total hazard construction method.  One can first
generate independent  standard
uniform  variables $U_i$ and set 
 $E_i=-\ln(1-U_i)$, $i=1,\ldots,n$, then
find default times one by one as follows:
To find the 
first default time $\tau^1$ and the corresponding name $j_1$, set 
$$j_1 = {\rm argmin}_{j=1,\ldots,n} \left\{t_j>0: \int_0^{t_j} \lambda_j(s)ds \geq
E_j\right\}$$
and $\tau^1=t_{j_1}$ and $J_1=\{j_1\}$, where $\lambda_j(s)$
are unconditional hazard rates  of names $j$ at time $s$.  
To find the $k$th default time $\tau^k$ and the corresponding name $j_k$ 
for $k\geq 2$, set
$$j_k = {\rm argmin}_{j \not\in J_{k-1}}
\left\{t_j>\tau_{j_{k-1}}: \int_0^{t_j} \lambda_j(s|t_{J_{k-1}})ds \geq
E_j\right\}$$ 
and $\tau^k=t_{j_k}$ and $J_k=J_{k-1}\cup \{j_k\}$.

We suggest a copula contagion mixture model which covers both copula model and contagion model as special cases. Specifically, we assume that the intensity processes
$\lambda_i$ may depend on default times of other names and standard
uniform variables $F_i(\tau_i)$ have a joint distribution $C$.
This is a natural generalization of pure copula models and pure contagion models. One can easily generate default times with the total hazard construction method. The only difference with the pure contagion model  
is that one 
generates  standard uniform variables $U_i$ from 
a given  copula $C$, not necessarily from the product copula 
which corresponds to the pure contagion model. 
The key advantage of this new mixture model is  that,
instead of studying three well known models in isolation, we can  
explore their interaction  and their joint pricing impact on CDOs and basket CDSs.

We now impose some  structure to the intensity processes. To simplify the
notation and highlight the key point, we assume a homogeneous portfolio.
The discussion is the same 
for general heterogeneous intensity processes 
except the expression is more complicated.  
We assume the intensity processes have the following structure
\begin{equation}
\lambda_i(t) = a\left(1+ \sum_{j=1, j\ne i}^n ce^{-d(t-\tau_j)}1_{\{\tau_j\leq t\}}\right),
\quad i=1,\ldots,n, \label{lambdai}
\end{equation}
where $a, c, d$ are positive constants.  
(These parameters can be deterministic  functions of $t$ or even some stochastic processes, the discussion is essentially the same, see  Zheng
and Jiang (2009).) 
$a$ is the unconditional default intensity, $c$ is the contagion rate, and
$d$ is the exponential decay rate. 
When $d=0$  we may introduce the default state space and use the Markov Chain to study the joint distribution of default times. Apart from this extreme case the intensity processes 
(\ref{lambdai}) are non-Markov.

For homogeneous intensity processes
(\ref{lambdai}) without exponential decay ($d=0$)
we can simplify the total hazard construction method. This is because we only need to know the number of defaults at time $t$ but not the identities of names which have defaulted. We can generate
  $\tau^k$  as follows: 

\medskip\noindent
{\it Step 1.} Generate correlated
standard uniform variables $U_i$, $i=1,\ldots,n$,
from the copula $C$.\\
{\it Step 2.} Set $E_i=-\ln(1-U_i)$, $i=1,\ldots,n$,
and sort $E_i$ in increasing order to get $E_i^*$ with $E_1^*< E_2^*< \ldots < E_n^*$.\\
{\it Step 3.} Find ordered default times $\tau^k$ by setting
\begin{equation} \tau^1={E_1^*\over a}, \quad
\tau^k=\tau^{k-1} + {E_k^*-E_{k-1}^*\over a(1+(k-1)c)}
\quad\mbox{ for }\quad k=2,\ldots,n. \label{tauk0}
\end{equation} 
\medskip

The density function of the 
$k$th  default time $\tau^k$ is given by (for $d=0$) 
$$  f_{\tau^k}(t)=\sum_{j=0}^{k-1}
\alpha_{k,j}a e^{-\beta_j at}$$
where $\beta_j = (n-j)(1+jc)$ and $\alpha_{k,j}$ are constants depending
on $c,k,j,n$ and have explicit expressions. For example, 
$f_{\tau^1}(t)=nae^{-nat}$ which shows that the
contagion has no influence on the first default time, and
$$
f_{\tau^2}(t)=\left\{\begin{array}{ll}
{n(n-1)(1+c)a\over (1+(1-n)c)} ( -e^{-nat} + e^{-(n-1)(1+c)at}),
& \mbox{ if } c\ne 1/(n-1)\\
(na)^2 te^{-nat}, & \mbox{ if } c= 1/(n-1) 
\end{array} \right.
$$
which implies that the
contagion affects the second and all subsequent default times.
We can then derive the analytic pricing formulas
for  basket CDSs and CDOs, see  Zheng and Jiang (2009).

For general homogeneous intensity processes (\ref{lambdai}) 
 we cannot use (\ref{tauk0}) to generate ordered default times. The computation is slightly more involved. Steps 1 to 2 are the same and so is the first default time
$\tau^1$. Assume ordered default times $\tau^1,\ldots,\tau^{k-1}$ have 
already been generated for some $k\geq 2$. Now we want to generate 
$\tau^{k}$. Let $\tau^{k-1}\leq t\leq \tau^{k}$. The total hazard accumulated by name $k$ at time $t$ is 
$$ \int_0^t \lambda_{k}(s)ds
=\sum_{j=1}^{k-1}\int_{\tau^{j-1}}^{\tau^j}
a\left(1+\sum_{i=1}^{j-1} ce^{-d(s-\tau^i)}\right) ds 
+ \int_{\tau^{k-1}}^t a\left(1+ \sum_{i=1}^{k-1} 
ce^{-d(s-\tau^i)}\right) ds$$
where $\tau^0=0$ and $\sum_{i=1}^0 = 0$ by convention. 
Simplifying the above expression we get
$$  \int_0^t \lambda_{k}(s)ds=
at + {ac\over d} \sum_{i=1}^{k-1} \left(1-e^{-d(t-\tau^i)}\right).$$
The $\tau^{k}$ is determined by the relation
$\int_0^{\tau^{k}} \lambda_{k}(s)ds =E_{k}^*$. 
Define 
$$ F_{k}(t) := 
at + {ac\over d} \sum_{i=1}^{k-1} \left(1-e^{-d(t-\tau^i)}\right) -E_{k}^*.$$
Then $\tau^{k}$ is a root of nonlinear equation $F_{k}(t)=0$. 
Since $F_{k}'(t)>0$ and $F_{k}''(t)<0$ function $F_{k}$ is strictly increasing and strictly concave. Observe also that from $F_{k-1}(\tau^{k-1})=0$
we have 
$$F_{k}(\tau^{k-1})= a\tau^{k-1}+ {ac\over d} \sum_{i=1}^{k-1} 
\left(1-e^{-d(\tau^{k-1}-\tau^i)}\right) -E_{k}^*=E_{k-1}^*-E_{k}^*< 0$$
and $F_{k}(\infty)=\infty$. There is a unique root of equation 
$F_{k}(t)=0$ on the interval $[\tau^{k-1},\infty)$. The special structure of function $F_{k}$ guarantees that
the Newton algorithm with an initial iterating point $\tau^{k-1}$
converges quadratically to the root $\tau^{k}$. 
We can now summarize Sept 3 
in the presence of exponential decay rate $d>0$ as follows.

\medskip\noindent
{\it Sept 3$'$.} Set $\tau^1=E_1^*/a$ and find the $k$th default time
$\tau^k$ by solving numerically  
the equation $F_k(t)=0$
with the Newton algorithm and the initial iterating point 
$\tau^{k-1}$ for $k=2,\ldots,n$. 
\medskip

We now discuss the impact of exponential decay rate $d$ on ordered default times $\tau^k$. From
$ F_k'(t)=a+ac\sum_{i=1}^{k-1}e^{-d(t-\tau^i)}$
we know that $F_k'(t)$ is a strictly decreasing function of $d$ for
$t>\tau^{k-1}$. If $d=0$ we have $F_k'(t)=a+ac(k-1)$ and $F_k$ is a linear function 
$$F_k(t)=E_{k-1}^* - E_k^*+ (a+ac(k-1))(t-\tau^{k-1}).$$
The $k$th default time $\tau^k$ is given by  (\ref{tauk0})
as expected. If
$d=\infty$ we have $F_k'(t)=a$ and $F_k$ is again a linear function 
$$ F_k(t)=E_{k-1}^* - E_k^*+ a(t-\tau^{k-1}).$$
The $k$th default time  is given by
$ \tau^k=\tau^{k-1}+(E_{k}^* - E_{k-1}^*)/a$, or equivalently,
$\tau^k=E_k^*/a$,
which corresponds to the case when there is no contagion effect.
For any other $d$ the $k$th default time $\tau^k$ lies between these two extreme cases. We conclude that the smaller the exponential decay rate, 
the stronger the contagion effect and the sooner the ordered default times, which makes CDO and basket CDS riskier and demands higher spreads.

\section{Numerical Tests}
We can now value the basket CDS and CDO with the copula contagion mixture model.
We assume homogeneous intensity processes 
(\ref{lambdai}) to simplify the computation, but the same method can be applied to general intensity processes. For both basket CDS and CDO we assume that $T$ is the maturity of the contract,
$t_1<t_2\ldots<t_N$ are swap rate payment dates, $t_0=0$ is the initial time and $t_N=T$ is the terminal time,  $R$
is the recovery rate, $r$ is the riskless interest rate, and
$B(t)=e^{-rt}$ is the discount factor at time $t$. 

To price basket CDS we assume
 $S_k$ is the annualized $k$th default swap
rate. 
The expected value of the contingent leg at time 0 is equal to 
$$ \E[(1-R)B(\tau^k)1_{\{\tau^k\leq T\}}]
$$
and that of the fee leg with accrued interest is equal to
$$ S_k \E\left[\sum_{i=1}^N \left((t_i-t_{i-1}) B(t_i)
1_{\{\tau^k>t_i\}}
+ (\tau^k-t_{i-1})B(\tau^k)
1_{\{t_{i-1}<\tau^k\leq t_i\}}\right)\right].$$
We can easily find the swap rate $S_k$ with the Monte Carlo method by generating ordered default times $\tau^k$.

To price CDO we assume 
$k_l$, $l=0,\ldots, M-1$, are attachment points of tranches $l$
with $0=k_0<k_1<\ldots<k_M=1$, 
$\Delta k_l = k_{l}-k_{l-1}$ are tranche sizes for $l=1,\ldots,M$,
the cumulative percentage portfolio loss at time $t$ is given by 
$$ L(t) = \sum_{k=1}^{n} {k\over n} 1_{\{\tau^k\leq t<\tau^{k+1}\}} $$
with $\tau^0=0$ and $\tau^{n+1}=\infty$, the cumulative tranche $l$ 
loss at time $t$ is given by
$$ L_{l}(t)=(L(t)-k_{l-1})1_{\{k_{l-1}\leq L(t)\leq k_l\}}
+\Delta k_l 1_{\{L(t)>k_l\}}.$$
Assume $S_l$ is the swap rate of tranche $l$. 
The expected  value of the contingent leg for 
tranche $l$ loss at time 0 is given by  (note $L_{l}(0)=0$)
$$
\E\left[\sum_{i=1}^N B(t_i)(L_l(t_i) - L_l(t_{i-1}))\right]
$$
and that of the fee leg for tranche  $l$ is 
$$ S_l\E\left[\sum_{i=1}^N (t_i-t_{i-1}) 
B(t_i)(\Delta k_l - L_l(t_i))\right].
$$
We can again easily find the swap rate $S_l$ with the Monte Carlo method. 

To generate ordered default times we must first generate correlated standard
uniform variables $U_i$, $i=1,\ldots,n$. We use three different copulas to
generate $U_i$.  The first one is the product copula and 
$U_i$ are simply independent standard uniform variables.
The second one is the exponential copula and $U_i$ are generated
as follows: first generate $n+1$ independent exponential variables $T_0,T_1\ldots,T_n$,
where $T_0$ has parameter $c_0$ and $T_1,\ldots,T_n$ have parameter $c_1$,
then set $S_i=\min(T_0,T_i)$, 
and finally define $U_i=1-\exp(-(c_0+c_i)S_i)$, $i=1,\ldots,n$. 
This is the simplest exponential copula which models simultaneous jumps as
well as individual jumps, 
see Giesecke (2003) for more details on exponential copulas and 
Xu and Zheng (2009) for their applications in modelling portfolio asset price
processes. The third model is the Gaussian
copula and $U_i$ are generated as follows:
first generate $n+1$ independent standard normal variables
$Z, Z_1,\ldots,Z_n$, then set $X_i=\rho Z+
\sqrt{1-\rho^2}Z_i$, $i=1,\ldots,n$, and finally define
$U_i=\Phi(X_i)$, $i=1,\ldots,n$. This is the most popular model used in 
financial institutions for pricing portfolio derivatives.

We have used the following data in numerical tests:
number of names $n=40$, riskless interest rate $r=0.05$, time to maturity $T=3$, number of payments $N=6$ with equally spaced time intervals, 
unconditional intensity rate $a=0.01$, recovery rate $R=0.5$, 
exponential decay rate $d=0$, and number of simulations is 1 million.

\begin{table} \label{table0}
\begin{center}
\begin{tabular}{|c|c|ccc|} \hline
$\rho/c$ & tranche & 0.0 & 0.3 & 3.0\\ \hline
0.0 & 0-0.15 & 0.0740 & 0.0890 & 0.2360\\
& 0.15-0.3 & 0.0000 & 0.0003 & 0.1052\\
& 0.3-1 & 0.0000 & 0.0000 & 0.0199\\
\hline
0.5 & 0-0.15 &0.0682  & 0.0843&0.1553\\
&0.15-0.3& 0.0042 & 0.0164 & 0.1020\\
& 0.3-1& 0.0001 &0.0022& 0.0596\\
\hline
0.9& 0-0.15&0.0326&0.0373&0.0488\\
&0.15-0.3&0.0147&0.0242&0.0439\\
&0.3-1& 0.0044 &0.0157&0.0405\\
\hline
\end{tabular}
\end{center}
\caption{CDO rates with the Gaussian copula contagion mixture model.}
\end{table}

Table~1 lists CDO rates computed 
with the  Gaussian copula contagion mixture model
with  different  $c$ and  $\rho$. 
We can see that swap rates increase if $c$ increases, which is expected as higher $c$ causes higher contagion and more defaults. $c=0$ corresponds to the Gaussian factor model. As $\rho$ increases swap rates for equity tranche decrease while those for mezzanine and senior tranches increase, a well known fact. $\rho=0$ corresponds to the pure contagion model 
(or the product copula contagion mixture model)
and we see $c$ has huge impact on swap rates for mezzanine and senior tranches. When both $c$ and $\rho$ are positive, we see swap rates for senior tranche are greater than those with the pure contagion model ($\rho=0$) 
and the pure factor  model ($c=0$). It is interesting to note that swap rates for mezzanine tranche  decrease as $\rho$ increases
when $c=3$, an opposite phenomenon to the case when $c=0$. 
This is not surprising 
because when $c=3$ the default intensity increases quickly for surviving names and many more names are likely to default, in other words, the mezzanine tranche behaves increasingly like the equity tranche, and therefore as $\rho$ increases the corresponding swap rates actually decrease. For the same reason the senior tranche behaves increasingly like the mezzanine tranche and its swap rates increase and then decrease as $\rho$ increases. We have also done numerical tests for $a=0.1$ and found that all tranches behave like the equity tranche and swap rates decrease as $\rho$ increases even when $c=0.3$      

Table~1 may shed some light on the cause of  recent financial crisis.  
Before the full scale credit crunch, 
the housing and stock  markets were  booming, the  credit
was cheaply and easily available, few individuals and companies defaulted,
and default rates from rating agencies were very low. 
Portfolio credit derivatives such as synthetic CDOs were in high demand. The Gaussian factor model (corresponding to $c=0$ in the table) was the most popular model used in financial institutions to price these securities. Table~1 shows that for the senior tranche (0.3-1) the risk is almost negligible
for $\rho=0.5$, and is still very small even for unlikely $\rho=0.9$. 
It seemed that the underwriter of CDS for CDO senior tranches could make huge profit from
premium fees with little risk,  almost like ``free lunch with vanishing risk''.
However, when there is contagion, which is the case for synthetic CDOs (the
actual loss can be many times over the nominal loss), the risk for the senior
tranche is much higher even in good economy ($a=0.01$). When $c=3$ and $\rho=0.5$, the 
swap rate for the senior tranche is 0.0596, in sharp contrast to 0.0001 when $c=0$ and $\rho=0.5$. Mis-pricing of synthetic CDOs could be one of the causes
which led to the financial crisis of underwriters of CDS for these
synthetic portfolio credit derivatives.

\begin{table} \label{t2}
\begin{center}
\begin{tabular}{|c|ccc|ccc|ccc|} \hline
 $c$ &  &0&&& 0.3 &&& 3&\\ \hline
 $k$ & ProdC  & ExpC & GausC & ProdC  & ExpC & GausC & ProdC  
 & ExpC & GausC\\ \hline
 1&0.2024&0.1575&0.1153&0.2024&0.1575&0.1153&0.2024&0.1575&0.1153\\
2&0.0634&0.0697&0.0508&0.0769&0.0811&0.0573&0.1401&0.1249&0.0855\\
5&0.0010&0.0026&0.0105&0.0052&0.0104&0.0197&0.0836&0.0866&0.0620\\
10&0.0000&0.0000&0.0014&0.0000&0.0001&0.0072&0.0486&0.0582&0.0492\\
20&0.0000&0.0000&0.0000&0.0000&0.0000&0.0016&0.0163&0.0263&0.0369\\
30&0.0000&0.0000&0.0000&0.0000&0.0000&0.0003&0.0024&0.0061&0.0274\\ \hline
0--0.15 & 0.0740 & 0.0742&0.0682 & 0.0890&0.0923&0.0843 & 0.2360&0.2218&0.1553\\
0.15--0.3 & 0.0000 &0.0000&0.0042 & 0.0003&0.0011&0.0164 & 0.1052&0.1246&0.1020\\
0.3--1 & 0.0000&0.0000&0.0001 & 0.0000&0.0000&0.0022 & 0.0199&0.0314&0.0596\\
\hline
\end{tabular}
\end{center}
\caption{Comparison of basket CDS and CDO
rates with the  product copula contagion
mixture model (ProdC),  
the exponential copula contagion mixture model (ExpC)
 with $c_0=0.01$ and $c_1=0.1$, and
the Gaussian copula contagion mixture model (GausC) with $\rho=0.5$.}
\end{table}

Table~2 lists basket CDS and CDO rates computed with
three copula contagion mixture models.
The copulas used are 
the product copula, the exponential copula
with $c_0=0.01$ and
$c_1=0.1$ (individual jumps is much more likely than  a systematic jump), and the  Gaussian copula with $\rho=0.5$. 
The results for basket CDS are mixed with
no single model dominating the others in pricing. Contagion has no influence
to the 1st default CDS rate and the product copula produces the highest rate.
When there is no contagion ($c=0$) or low contagion ($c=0.3$) 
the Gaussian copula dominates the 
swap rates for all $k$ but the first few. 
When there is high contagion ($c=3$) the results are more mixed with 
the Gaussian copula 
dominating for large $k$ and the other copulas for 
small $k$. The results for CDO are also mixed. 
For equity tranche the product and exponential copulas 
produce similar
rates which are higher than those from the Gaussian copula. For mezzanine and senior
tranches the Gaussian copula gives much higher rates than the other two copulas
do except when contagion is high ($c=3$) and the rates from the other two copulas
are also increased significantly. The difference between swap rates 
using different copula contagion mixture models is   
substantial.

\begin{table} \label{t4}
\begin{center}
\begin{tabular}{|c|ccccc|} \hline
 $k$ & $d=0$&$d=1$&$d=10$&$d=100$&$d=\infty$\\ \hline
 1&0.1153&0.1153&0.1153&0.1153&0.1153\\
 2&0.0855&0.0761&0.0564&0.0514&0.0508\\
5&0.0620&0.0482&0.0175&0.0111&0.0105\\
10&0.0492&0.0348&0.0053&0.0017&0.0014\\
20&0.0369&0.0230&0.0008&0.0001&0.0000\\
30&0.0274&0.0137&0.0001&0.0000&0.0000\\ \hline
 0--0.15&0.1553&0.1323&0.0810&0.0696&0.0682\\
 0.15--0.3&0.1020&0.0727&0.0127&0.0048&0.0042\\
0.3--1&0.0596&0.0328&0.0012&0.0002&0.0001\\ \hline
\end{tabular}
\end{center}
\caption{Comparison of basket CDS and CDO
rates with exponential decay 
 Gaussian  copula contagion
mixture model, data used are $a=0.01$, $\rho=0.5$, and $c=3$.}
\end{table}     

Table~3 lists basket CDS and CDO
rates with the exponential decay  Gaussian copula contagion mixture model. The data used are $a=0.01$, $c=3$, $\rho=0.5$, 
and different decay rates. $d=0$ corresponds to the 
 Gaussian copula contagion mixture model without decay and $d=\infty$ to the case without contagion effect. It is clear that as $d$ increases, basket CDS and CDO
rates decrease. The exponential decay has much greater impact to the $k$th default rates for larger $k$ than for smaller  $k$. 
The same phenomenon is observed for CDO rates, that is,
 the exponential decay has much greater impact to senior tranche rates
  than to junior tranche rates. 
Basket CDS and CDO rates are highly sensitive to exponential decay rates $d$, which requires an accurate estimation of $d$ in calibration if one is to use it in pricing.

We have done some numerical tests
on pricing of CDOs and basket CDSs in the presence of 
counterparty risk. 
Assume the intensity processes (\ref{lambdai}) for underlying 
names in the portfolio and assume  the intensity process
of default time $\tau^B$ of the counterparty is given by  
$$
 \lambda_B(t)=a_B\left(1+\sum_{i=1}^n c_B 1_{\{\tau^i\leq t\}}\right) 
$$
 where $a_B$ is the unconditional default intensity and $c_B$ the
 contagion rate. 
Note that  the hazard rate process $\lambda_B$
 of the counterparty is influenced by
defaults of names in the portfolio, but not vice versa.  This follows the
observation in Leung and Kwok (2005) and Yu (2007) that the contagion of
the counterparty  on underlying names does not affect CDS pricing.

To price basket CDS  we only need to compute the expected
value of the contingent leg and the fee leg at time 0, given respectively by 
$$ \E\left[(1-R)B(\tau^k)1_{\{\tau^k\leq T, \tau^B\geq \tau^k\}}\right]$$
and 
$$ S_k \E\left[\sum_{i=1}^N \left( (t_i-t_{i-1}) B(t_i)
1_{\{\tau^k>t_i,\tau^B>t_i\}}
+ (\tau^k-t_{i-1})B(\tau^k)
1_{\{t_{i-1}<\tau^k\leq t_i, \tau^B>\tau^k\}}\right)\right].$$

Similarly to price  CDO tranche $l$
we only need to compute the expected
value of the contingent leg and the fee leg at time 0, given respectively by $$
\E\left[\sum_{i=1}^N B(t_i)(L_l(t_i) - L_l(t_{i-1}))
1_{\{\tau^B>t_i\}}\right]
$$
and 
$$ S_l\E\left[\sum_{i=1}^N (t_i-t_{i-1}) B(t_i)(\Delta k_l - L_l(t_i))
1_{\{\tau^B>t_i\}}\right]$$
We can easily find swap rates with the total hazard construction  method by generating ordered default times $\tau^k$ and counterparty default time $\tau^B$.

\begin{table} \label{t6}
\begin{center}
\begin{tabular}{|c|c|cc|cc|cc|} \hline
 & $c$ & 0.0 & & 0.3 &  &3.0 &\\ \hline
$\rho$ & tranche & GausC &GausCCR & GausC & GausCCR &GausC &GausCCR\\ \hline
0.0 & 0-0.15 & 0.0740 & 0.0740 & 0.0890&0.0889 & 0.2360&0.2347\\
& 0.15-0.3 & 0.0000 &0.0000 & 0.0003&0.0003 & 0.1052&0.1027\\
& 0.3-1 & 0.0000&0.0000 & 0.0000&0.0000 & 0.0199&0.0188\\
\hline
0.5 & 0-0.15 &0.0682&0.0680  & 0.0843&0.0841&0.1553&0.1521\\
&0.15-0.3& 0.0042&0.0040 & 0.0164&0.0160 & 0.1020&0.0968\\
& 0.3-1& 0.0001&0.0001 &0.0022&0.0020& 0.0596&0.0500\\
\hline
0.9& 0-0.15&0.0326&0.0326&0.0373&0.0364&0.0488&0.0421\\
&0.15-0.3&0.0147&0.0144&0.0242&0.0232&0.0439&0.0355\\
&0.3-1& 0.0044&0.0040 &0.0157&0.0137&0.0405&0.0291\\
\hline
\end{tabular}
\end{center}
\caption{CDO rates with the Gaussian copula contagion mixture 
model (GausC) and the same model with counterparty risk (GausCCR).}
\end{table}

Table~4 lists the swap rates of all tranches with and without counterparty risk. We have used the data $a_B=a/10=0.001$ and $c_B=c$.
The counterparty is much less likely to default than those names in the portfolio. However, defaults of names in the portfolio increase the default intensity of the counterparty. It can be observed that
the equity tranche is least affected by the counterparty risk while the senior tranche is most affected. This is expected as the counterparty is much
more likely to default due to the contagion effect from 
defaults of names in equity and mezzanine tranches,  and therefore
 the senior tranche investors require higher compensation for 
 increased counterparty risk.

\section{Conclusions}
In this paper we have suggested a general exponential decay
 copula contagion mixture model 
which unifies the factor model, copula model, and contagion model. The key advantage is that one can investigate the interaction of these models and its pricing impact on basket CDS and CDO. The ordered default times can be easily generated with the total hazard construction method. We have done some numerical tests and
compared basket CDS and CDO rates with three  copula (product, exponential, and Gaussian) contagion mixture models and found  that
there is no model dominating the others in pricing although one model may affect much greatly the pricing of some parts of basket CDS and CDO 
than the other models do. The difference of swap rates computed with different models can be substantial. The exponential decay rate has great impact on senior tranche rates and $k$th default rates for large~$k$. We have 
also compared pricing results when there is contagion counterparty risk.
Our conclusion is that one has to be cautious in pricing basket CDS and CDO when a particular model is used  
as different models may greatly influence 
the portfolio loss distribution and can significantly affect the resulting swap rates. 
We should not put all blames on the ``misplaced reliance on sophisticated maths'' for the recent financial crisis, see Cookson (2009).
No model is best for all purposes. 
Stress test and other risk control procedures should be in place to withstand the potential loss due to the wrong choice of models.

\bigskip\noindent{\bf Acknowledgement.} 
The author thanks Duanpeng Wang  
 for the help in numerical tests with C++.
The author also thanks
 the London Mathematical Society for its Scheme 4 grant support
  (Grant 4818).

\bigbreak\noindent{\Large\bf References.}
\begin{enumerate}
\item \label{cookson} 
Cookson, C., Count us in, maths experts tell regulator,
{\it Financial Times}, June 10, 2009.
\item \label{andersen}
Andersen, L., Sidenius, J., and Basu, S., All your hedges in one basket,
{\it Risk}, 67-72, November 2003. 
\item Das, S.R., Duffie, D., Kapadia, N., Saita, L., Common failings: how
corporate defaults are correlated, {\it J. Finance} 62, 2007.
\item \label{davislo}
Davis, M.H.A. and Lo, V., Infectious defaults, {\it Quantitative Finance}
1, 305-308, 2001.
\item\label{fb04}
 Frey, R.,  Backhaus, J.: Portfolio credit risk models with interacting
default intensities: a Markovian approach. Working paper. (2004)\\
{\tt http:$\backslash\backslash$www.math.uni-leipzig.de$\backslash$\%7Efrey$\backslash$interacting-intensities-final.pdf}
\item \label{giesecke}
Giesecke, K., A simple exponential model for dependent defaults, {\it J.
Fixed Income} 13, 74-83, 2003.
\item \label{glasserman}
Glasserman, P., Tail approximations for portfolio credit risk, {\it
J. Derivatives}, 24-42, Winter, 2004.
\item\label{hr06}
 Herbertsson, A., Rootzen, H.: Pricing $k$th-to-default swaps under
default contagion: the matrix-analytic approach. Working paper (2006)\\
{\tt http:$\backslash\backslash$www.math.chalmers.se$\backslash\sim$rootzen$\backslash$papers$\backslash$Herbertsson$\_$Rootzen$\_$2006.pdf}
\item \label{hullwhite}
Hull, J. and White, A., Valuation of a CDO and an $n$th to default
CDS without Monte Carlo simulation, {\it J. Derivatives} 12, 8-23, Winter
2004.
\item \label{jarrowyu}
Jarrow, R. and Yu, F., Counterparty risk and the pricing of defaultable
securities, {\it J. Finance} 53, 2225-2243, 2001.
\item Leung, S.Y. and Kwok, Y.K., Credit default swap valuation with counterparty
risk, {\it The Kyoto Economic Review} 74, 25-45, 2005.
\item Li, D.X., On default correlation: a copula function approach, working
paper, 2000. \\
{\tt http:$\backslash\backslash$www.defaultrisk.com$\backslash$pp$\_$corr$\_$05.htm}
\item Mikosch, T., Copulas: tales and facts, {\it Extremes} 9, 3-20, 2006.
\item Xu, G. and H. Zheng, Approximate basket options valuation for a jump-diffusion
model, {\it Insurance: Mathematics and Economics}, article in press, 2009. 
\item \label{yu}
Yu, F., Correlated defaults in intensity-based models,
{\it Mathematical Finance} 17, 155-173, 2007. 
\item \label{zheng}
Zheng, H., Efficient hybrid methods for portfolio credit derivatives,
{\it Quantitative Finance} 6, 349-357, 2006.
\item \label{zhengjiang}
Zheng, H. and L. Jiang, 
Basket CDS Pricing with Interacting Intensities, 
{\it Finance and Stochastics} 13, 445-469, 2009. 

\end{enumerate}

\end{document}